# An Integrated Framework for the Heterogeneous Spatio-Spectral-Temporal Fusion of Remote Sensing Images

Menghui Jiang [1], Huanfeng Shen [1, 3]*, Jie Li [2]*, Liangpei Zhang [3,4]

[1] School of Resource and Environmental Science, Wuhan University, P. R. China

[2] School of Geodesy and Geomatics, Wuhan University, P. R. China

[3] Collaborative Innovation Center of Geospatial Technology, Wuhan University, P. R. China

[4] State Key Laboratory of Information Engineering, Survey Mapping and Remote Sensing, Wuhan University

## Abstract

Image fusion technology is widely used to fuse the complementary information between multi-source remote sensing images. Inspired by the frontier of deep learning, this paper first proposes a heterogeneous-integrated framework based on a novel deep residual cycle GAN. The proposed network consists of a forward fusion part and a backward degeneration feedback part. The forward part generates the desired fusion result from the various observations; the backward degeneration feedback part considers the imaging degradation process and regenerates the observations inversely from the fusion result. The proposed network can effectively fuse not only the homogeneous but also the heterogeneous information. In addition, for the first time, a heterogeneous-integrated fusion framework is proposed to simultaneously merge the complementary heterogeneous spatial, spectral and temporal information of multi-source heterogeneous observations. The proposed heterogeneous-integrated framework also provides a uniform mode that can complete various fusion tasks, including heterogeneous spatio-spectral fusion, spatio-temporal fusion, and heterogeneous spatio-spectral-temporal fusion. Experiments are conducted for two challenging scenarios of land cover changes and thick cloud coverage. Images from many remote sensing satellites, including MODIS, Landsat-8, Sentinel-1, and Sentinel-2, are utilized in the experiments. Both qualitative and quantitative evaluations confirm the effectiveness of the proposed method.

# 1. Introduction

Due to the hardware limitations, remote sensing system imaging has to trade-off among resolution indexes, including temporal, spatial, and spectral resolutions (Zhang, 2004). Remote sensing image fusion is an effective way to fuse the complementary information between multi-source observations, which has been widely concerned and developed (Pohl and van Genderen, 1998; Sirguey et al., 2008; Martha et al., 2012). To date, a variety of remote sensing image fusion methods have been proposed; according to the fusion aim, they can be divided into many categories, including spatio-spectral fusion, spatio-spectral fusion, spatio-spectral-temporal fusion (Shen et al., 2016).

Spatio-spectral fusion (Shen et al., 2019) aims at obtaining images of both high spatial and spectral resolutions by fusing the complementary rich spatial and rich spectral features between two images. It mainly includes panchromatic (PAN)/multispectral (MS) images fusion, PAN/ HS images fusion, and MS/HS images fusion. The existing spatio-spectral fusion methods can be broadly classified into four major branches (Meng et al., 2018): the component substitution (CS)-based methods (Carper et al., 2004; Choi et al., 2011), the multiresolution analysis (MRA)-based methods (Aiazzi et al., 2006; Shahdoosti et al., 2017), the variational model-based methods (Zhang et al., 2012; Jiang et al., 2014) and the deep learning (DL)-based methods (Wei et al., 2017; Jiang et al., 2020). The above-mentioned methods aim to fuse homogeneous optical images. In addition, scholars have also proposed some heterogeneous spatio-spectral fusion methods, namely Synthetic aperture radar (SAR)-optical images fusion, which uses the rich spatial features in SAR images to make up for the optical images with rich spectral information. Most of the existing SAR-optical fusion methods (Alparone et al., 2004; Chen et al., 2010) are transferred from the spatio-spectral fusion of optical images, are unsuitable for heterogeneous information transformation.

Spatio-temporal fusion (Wu et al., 2020) aims at obtaining images with both high spatial and temporal resolutions by fusing high spatial resolution (HR) images with a long revisit period and low spatial resolution (LR) images with a short revisit period. Spatio-temporal fusion methods can be broadly classified into four categories (Li et al., 2020): weight function-based methods (Gao et al., 2006; Cheng et al., 2017), unmixing-based methods (Zhukov et al., 1999; Huang et al., 2014), Bayesian-based methods (Li et al., 2013; Xue et al., 2017), and learning-based methods (Song et al., 2013; Li et al., 2020). Most of them can capture phenological changes, but they have a bottleneck that is hard to reflect the abrupt land cover changes, which is a common problem in spatio-temporal fusion under large spatial resolution gaps or severe weather conditions (such as thick cloud coverage).

The aforementioned spatio-spectral fusion and spatio-temporal fusion are dedicated to fusing only two of the spatial, temporal, and spectral resolution indexes. On this basis, the integrated fusion aiming at obtaining images of high spatial-spectral-temporal resolutions was first proposed by Shen (Shen et al., 2012). Huang (Huang et al., 2013) extended the model in (Shen et al., 2012) by exploring the relationship between the spatio-spectral and spatio-temporal fusion methods. Shen (Shen et al., 2016) proposed an integrated framework that thoroughly analyzed the spatial, spectral, and temporal relationships between the desired image and the multi-source remote sensing observations and constructed an integrated relationship model based on the maximum a posteriori (MAP) theory. Zhao (Zhao et al., 2018) exploited the high self-similarity in the spatial domain, the high-spectral correlation in the spectral domain, and the temporal changes to develop an integrated sparsity model. However, since the relationships between multi-source datasets are complex and difficult to describe accurately, the current studies on integrated fusion methods are limited; and they are all limited to homogeneous optical images without considering heterogeneous information.

To address these issues, for the first time, we propose a deep residual cycle generative adversarial network (GAN)-based heterogeneous-integrated fusion framework. The major contributions of this paper include:

1) We propose a novel deep residual cycle GAN, which consists of a forward fusion part and a backward degeneration feedback part. The proposed network can effectively fuse both homogeneous and heterogeneous information.

2) We are the first to propose a heterogeneous-integrated framework that can simultaneously merge the complementary heterogeneous spatial, spectral and temporal information between multi-sources optical and SAR images. In addition, the proposed heterogeneous-integrated framework provides a uniform mode that can deal with various tasks, including heterogeneous spatio-spectral fusion, spatio-temporal fusion, and heterogeneous spatio-spectral-temporal integrated fusion.

3) Three types of experiments are conducted for two challenging scenarios of land cover changes and thick cloud coverage. They are spatial resolution improvement experiments, thick cloud removal experiments, and simultaneous thick cloud removal and spatial resolution improvement experiments.

The remainder of the paper is organized as follows. Section II describes the proposed deep residual cycle GAN in detail. In Section III, the experiments and analysis on two challenging scenarios are presented. The conclusions and future prospects are reported in Section IV.

## 2. The Proposed Method

Before describing the proposed method in detail, some notations are introduced. $X \in \mathbb{R}^{M \times N \times B}$ denotes the desired HR MS image at the target time t1, where M, N, and B represent the width, the height, and the band number of the image respectively. $\tilde{X} \in \mathbb{R}^{m \times n \times B}$ denotes the observed LR MS image at t1. S=M/m=N/n is the spatial resolution ratio of the LR MS image to the HR MS image. $\hat{X} \in \mathbb{R}^{M \times N \times B}$ is the result of $\tilde{X}$ bicubic up-sampling to the same spatial size of $X$. $Y \in \mathbb{R}^{M \times N \times b}$ denotes the observed HR SAR image at t1, where b< B. $Z \in \mathbb{R}^{M \times N \times B}$ denotes the observed HR MS image at the auxiliary time t2. The relationships between the observations and the desired image can be formulated as:

$$\begin{cases} \tilde{X} = f_{spatial}(X) \\ Z = f_{temporal}(X) \\ Y = f_{heterogeneous}(X) \end{cases} \quad (1)$$

where $f_{spatial}(\cdot)$ denotes the spatial degradation relationship from $X$ to $\tilde{X}$, which is usually assumed to be a blurring and downsampling operation (Shen et al., 2019). $f_{temporal}(\cdot)$ denotes the temporal relationship from $X$ to $Z$, which is usually assumed to be a linear model (Fasbender et al., 2007, Huang et al., 2013). $f_{heterogeneous}(\cdot)$ denotes the heterogeneous spectral relationship between $X$ and $Y$, which is currently difficult to be expressed explicitly.

Fig. 1 displays the flowchart of the proposed method with the heterogenous spatio-spectral-temporal integrated fusion task as an example. The proposed deep residual cycle GAN is based on GAN (Goodfellow et al., 2014) with some improvements. In the network training in Fig. 1 (a), the network can be divided into a forward part and a backward degeneration feedback part. The forward part consists of a forward generator and a forward discriminator. The inputs of the forward generator for the heterogenous-integrated fusion task are t1 LR MS, t1 HR SAR, and t2 HR MS images. For the heterogeneous spatio-spectral fusion task, the inputs are t1 LR MS and t1 HR SAR images; for the spatio-temporal fusion, the inputs are t1 LR MS and t2 HR MS images. In the following, we describe the proposed network in detail only through the heterogenous-integrated fusion task. As shown in Fig. 1 (a), the output of the forward generator is the fused HR MS image at t1. It can be written as:

$$fusion = \boldsymbol{G}_F\left((\hat{X}, Y, Z); \Theta_F\right) \quad (2)$$

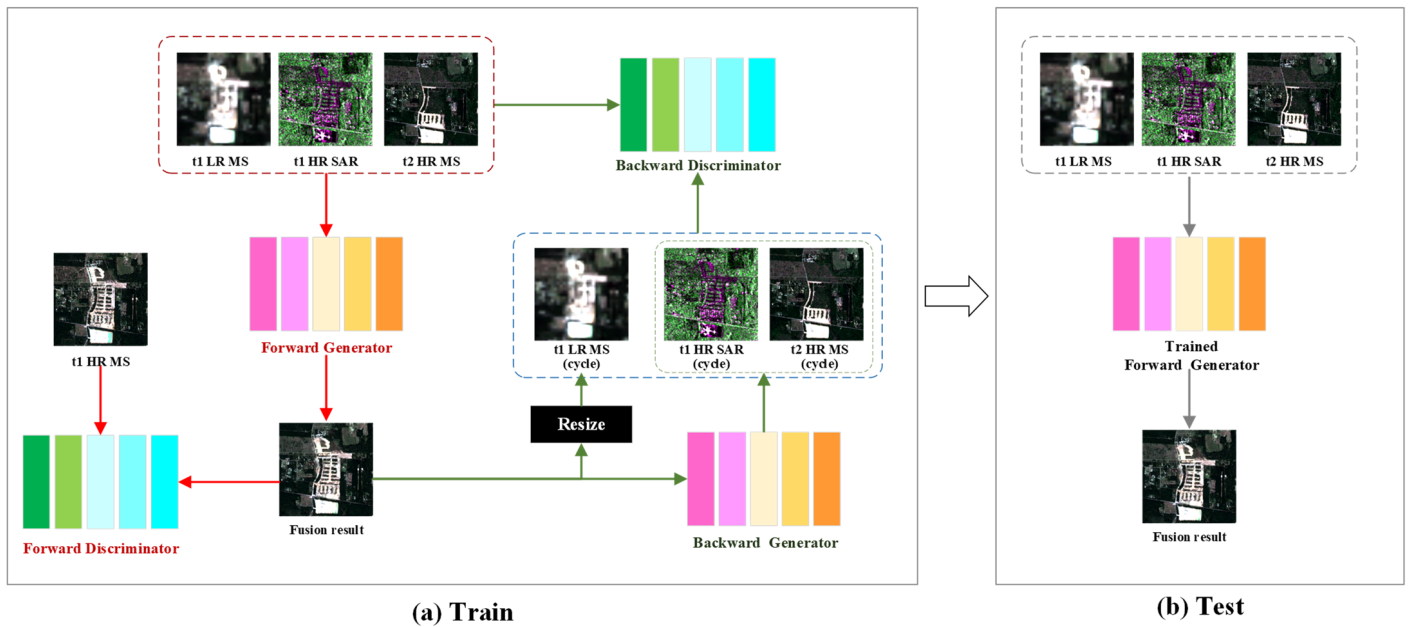

**Fig. 1.** Flowchart of the proposed method.

where *fusion* is the output of the forward generator $G_F(\cdot)$, $\Theta_F$ is the trainable parameter. The forward discriminator discriminates *fusion* and the label data, i.e. the t1 HR MS image.

The backward degeneration feedback part takes the degradation process of remote sensing imaging into account and reversely generates observation images from the fusion result. As shown in Fig. 1 (a), it consists of the 'Resize' branch, the backward generator, and the backward discriminator. Among them, the 'Resize' branch generates the t1 LR MS image from the fusion result according to the spatial degradation function $f_{spatial}(\cdot)$ in (1). Since the t2 LR MS image, which is usually necessary for estimating the temporal relationship model (Fasbender et al., 2007, Huang et al., 2013), is not utilized in the proposed method, we implicitly implement the temporal relationship function $f_{temporal}(\cdot)$ and the heterogeneous spectral relationship function $f_{heterogeneous}(\cdot)$ through the backward generator. They can be written as:

$$\hat{X}^* = resize(fusion) \tag{3}$$

$$Y^*, Z^* = G_B(fusion; \Theta_B) \tag{4}$$

where $resize(\cdot)$ represents the spatial downsampling operation. Note that in the case of thick cloud coverage, $resize(\cdot)$ represents the sequential operation of spatial downsampling and adding the cloud mask. $\hat{X}^*$ is the regenerated t1 LR MS image. $G_B(\cdot)$ and $\Theta_B$ denote the backward generator and the corresponding trainable parameter, $Y^*, Z^*$ denote the regenerated t1 HR SAR and t2 HR MS images. A cycle forms from the inputs of the forward generator to the outputs of the 'Resize' branch and backward generator. The backward discriminator discriminates $(\hat{X}, Y, Z)$ and $(\hat{X}^*, Y^*, Z^*)$. In the network testing in Fig. 1 (b), we input the observations into the trained forward generator, the output is our final fusion result.

## 2.1. Architectures of the proposed network

The proposed deep residual cycle GAN includes two generators and two discriminators. The two generators have the same network structure, and the two discriminators have the same network structure. The adopted generator network is similar to that of (Zhu et al., 2017), which consists of a feature extraction module, a feature coding module, a residual learning module, a feature decoding module, and a feature compression module. Fig. 2 shows the detailed structure of it. First, the input is fed into the feature extraction module, which is a 'Conv+BN+ReLU' block that composed of a convolution layer, a batch normalization layer, and a ReLU activation function layer. The convolution layer consists of 64 filters of 7×7×InC with stride 1, and InC denotes the channels of the network input. The feature encoding module down-samples the feature maps by convolutional layers with stride 2 (Springenberg et al., 2015), which increases the receptive field of features without increasing the kernel size of convolution or the network depth. As shown in Fig. 2, the feature encoding module consists of two 'Conv+BN+ReLU' blocks, in which the first convolution layer consists of 128 filters of 3×3×64, the second convolution layer consists of 256 filters of 3×3×128. The following residual learning module consists of multiple residual blocks. It utilizes the popular residual learning strategy (He et al., 2016), whose effectiveness has been verified in many tasks (Kiku et al., 2013; Timofte et al., 2014). As shown in Fig. 2, a residual block consists of a 'Conv+BN+ReLU' structure and a 'Conv+BN' structure, in which both convolution layers include 256 filters of 3×3×256. ⊕ denotes a pixel-by-pixel addition function. The feature decoding module has the opposite function to the feature encoding module. It gradually expands the feature map to the input size through deconvolution (Ronneberger et al., 2015). It composes of two 'DeConv+BN+ReLU' blocks. The first deconvolutional layer includes 128 filters of 3×3×256 with stride 2, and the second deconvolutional layer includes 64 filters of 3×3×128 with stride 2. The final feature compression module maps the features back to the image domain, it consists of a 'Conv+Tanh' block, in which the convolution layer includes OutC filters of 7×7×64, and OutC denotes the channels of the network output.

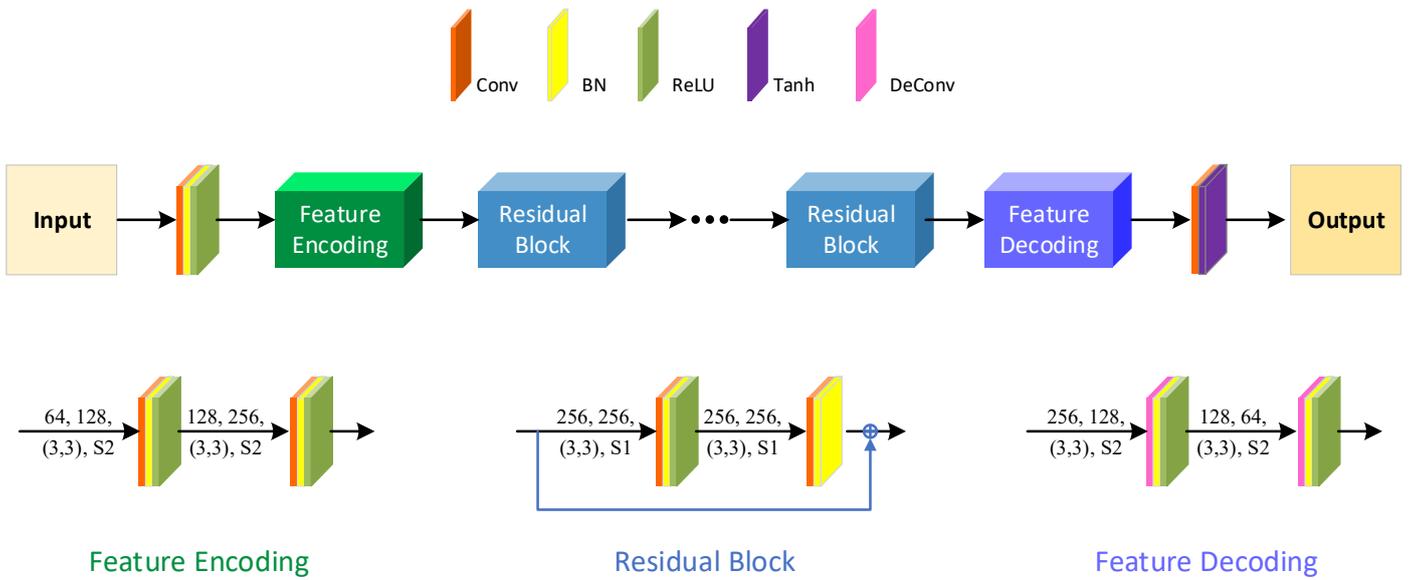

**Fig. 2** Structure of the proposed generators.

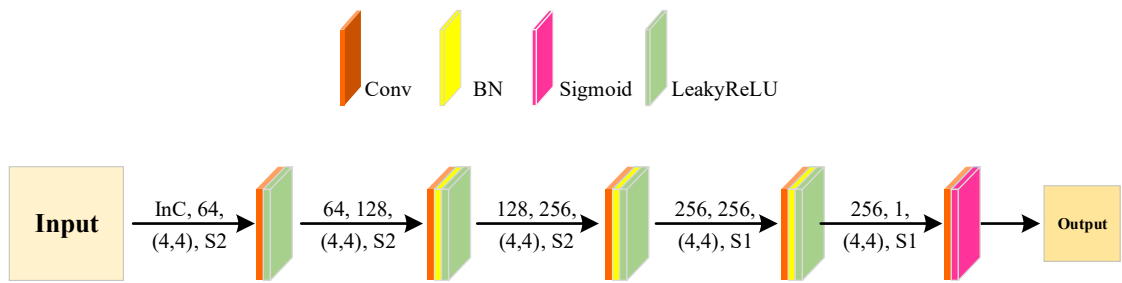

**Fig. 3** Structure of the proposed discriminators.

For the discriminator networks, we use PatchGAN (Isola et al. 2017), which classifies whether the overlapping image patches are real or fake. Fig. 3 shows the detailed structure of the proposed discriminator, which consists of one 'Conv+LeakyReLU' block, three 'Conv+BN+LeakyReLU' blocks, and a 'Conv+Sigmoid' block. The kernel size of all convolution layers is 4×4, and the stride of the first three convolution layers is 2, of the last two convolution layers is 1.

**2.2 Loss Function**

Consistent with the network structure, the loss function of the proposed network also includes two parts: the loss function of generator networks and the loss function of discriminator networks. The two generators are trained together with one loss function, which can be written as follows:

$$L_G = L_{adv} + \lambda L_{con} \tag{5}$$

where $L_G$ denotes the total loss of the generators and consists of two terms: $L_{adv}$ and $L_{con}$. $L_{adv}$ is the adversarial loss between generators and discriminators, which is defined as:

$$L_{adv} = \frac{1}{N}\sum_{n=1}^{N}\left\|D_F(fusion)-1\right\|_F^2 + \frac{1}{N}\sum_{n=1}^{N}\left\|D_B\left((\widehat{X}^*, Y^*, Z^*)\right)-1\right\|_F^2 \tag{6}$$

where the first term is the forward discriminator-related adversarial loss, and the second is the backward discriminator-related adversarial loss. $N$ denotes the number of patches in a batch. The MSE loss is empirically utilized in $L_{adv}$, and $\|\cdot\|_F$ is the matrix Frobenius norm.

$L_{con}$ in (5) is the content loss to ensure that the outputs of generators are close to the ground truths, and $\lambda$ is the adjustable parameter that balances $L_{adv}$ and $L_{con}$. Specifically, $L_{con}$ is defined as follows:

$$L_{con} = \lambda_1 * \frac{1}{N}\sum_{n=1}^{N}\left\|fusion-X\right\|_1 + \lambda_2 * \frac{1}{N}\sum_{n=1}^{N}\left\|(\widehat{X}^*, Y^*, Z^*)-(\widehat{X}, Y, Z)\right\|_1 \tag{7}$$

where the first term calculates the loss between the forward generator output and the ideal fusion result. The second term is the cyclic consistency between the inputs of the forward generator and the outputs of the 'Resize' branch and the backward generator. Since MAE loss is less sensitive to outliers than MSE loss, it is empirically used in the content loss, and $\|\cdot\|_1$ is the L1 norm.

The two discriminators are trained separately with separate loss functions. The forward discriminator distinguishes the forward generator output and the ideal fusion result, i. e. t1 HR MS image. It judges t1 HR MS image as true with label 1, and the output of the forward generator as fake with label 0. The loss function can be formulated as follows:

$$L_{D_F} = \frac{1}{2}\left(\frac{1}{N}\sum_{n=1}^{N}\|\boldsymbol{D_F}(fusion) - 0\|_F + \frac{1}{N}\sum_{n=1}^{N}\|\boldsymbol{D_F}(X) - 1\|_F\right) \qquad (8)$$

Similarly, the backward discriminator distinguishes the inputs of the forward generator and outputs of the 'Resize' branch and the backward generator. It judges the former as true with label 1, and the latter as fake with label 0. The loss function can be formulated as follows:

$$L_{D_B} = \frac{1}{2}\left(\frac{1}{N}\sum_{n=1}^{N}\|\boldsymbol{D_B}\left((\hat{X},Y,Z)\right) - 1\|_F + \frac{1}{N}\sum_{n=1}^{N}\|\boldsymbol{D_B}\left((\hat{X}^*,Y^*,Z^*)\right) - 0\|_F\right) \qquad (9)$$

In the network training, the trainable parameters of the generators and the discriminators are updated sequentially according to the corresponding loss functions.

## 3. Experiments

### 3.1. Comparison methods and quantitative evaluation indices

To verify the effectiveness of the proposed method, we conduct three types of experiments for two challenging problems of land cover changes and thick cloud coverage. They are spatial resolution improvement experiments, thick cloud removal experiments, and simultaneous thick cloud removal and spatial resolution improvement experiments. Datasets from multiple remote sensing satellites, including Moderate-resolution Imaging Spectroradiometer (MODIS), Sentinel-2 Multispectral Imager, Sentinel-1, and Lansat-8 Operational Land Imager (OLI), are utilized. Among them, the revisit period of MODIS is about one day; we use the red, green, and blue bands (i.e., band1, band4, and band3) in the MODO9GA product of MODIS, whose spatial resolution is 500m. The revisit period of the double Sentinel-2 satellites is five days. We use the red, green, and blue bands (i.e., B04, B03, and B02) in the Level-1C product of Sentinel-2, whose spatial resolution is 10m. The revisit period of the double Sentinel-1 satellites is six days. We use the VH and VV polarization bands of the Ground Range Detected (GRD) product of Strip Map Mode (SM) in Sentinel-1, whose resolution is 10m. The revisit period of the Landsat-8 is sixteen days. We use the red, green, and blue bands (i.e., B4, B3, and B2) of the Level-1 Precision Terra (L1TP) product of Landsat-8, whose spatial resolution is 30m.

Five mainstream fusion algorithms are selected for the comparison. Specifically, in the spatial resolution improvement experiments, ESRCNN (Shao et al., 2019) belongs to the spatio-temporal fusion strategy is selected to be the benchmark algorithm. In the cloud removal experiments, SAR-opt-cGAN (Grohnfeldt et al., 2018) and Simulation-Fusion GAN (Simu-Fus-GAN) (Gao et al., 2020) belong to the heterogeneous spatio-spectral fusion strategy are utilized. The modified version of STNLFFM (MSTNLFFM) (Shen et al., 2019) and the spatial–temporal–spectral deep convolutional neural network (STS-CNN) (Zhang et al., 2018) belong to the spatio-temporal fusion strategy are utilized. In the simultaneous thick cloud removal and spatial resolution improvement experiments, due to the lack of related researches, no comparison methods are selected, we focus on the internal comparison of our proposed method with three fusion strategies. Five representative indices are used to evaluate the performance of the fusion results quantitatively. They are the relative dimensionless global error in synthesis (ERGAS) (Vivone et al., 2014), the spectral angle mapper (SAM) (Vivone et al., 2014), the Q metric (Vivone et al., 2014), the peak-signal-to-noise ratio (PSNR), and the structural similarity index (SSIM) (Wang et al., 2004).

Table I
DATA SETS OF SPATIAL RESOLUTION IMPROVEMENT EXPERIMENTS

| | Sensor | Time | Resolution | Size (train) | Size (test) | Location (train) | Location (test) |
|---|---|---|---|---|---|---|---|
| Spatial Resolution Improvement | MODIS | 2017-10-29 | 500m | 128×106×3<br>60×60×3<br>124×130×3 | 60×60×3 | 95.62W, 30.23N<br>95.43W, 29.86N<br>95.77W, 29.41N | 95.78W, 29.85N |
| | Sentinel-1 SAR | 2017-10-28 | 10m | 6400×5300×2<br>3000×3000×2<br>6200×6500×3 | 3000×3000×2 | | |
| | Sentinel-2 MS | 2016-09-29 | 10m | 6400×5300×3<br>3000×3000×3<br>6400×5300×3 | 3000×3000×3 | | |
| | Sentinel-2 MS | 2017-10-29 | 10m | 6400×5300×3<br>3000×3000×3<br>6400×5300×3 | 3000×3000×3 | | |

## 3.2. Spatial resolution improvement experiments

### 3.2.1. Data sets for training and testing

In the spatial resolution improvement experiments, we improve the spatial resolution of the MODIS MS images at t1 from 500m to 10m with the help of the Sentinel-1 SAR image at t1 and Sentinel-2 MS image t2. Table I lists the data sets used for network training and testing. Note that in the network training and the network testing, to reduce the interference of many factors like registration, the t1 MODIS image is synthesized from the t1 Sentinel-2 image through bicubic downsampling. In table I, 2016-09-29 is t2, and the other three are t1. Three image pairs are utilized to generate the training patches, where the sizes of the MODIS images are 128×106, 60×60, and 124×130, of the others are 6400×5300, 3000×3000, 6200×6500. Their center locations are (95.62W, 30.23N), (95.43W, 29.86N), (95.77W, 29.41N). In total, 1984 patches of size 200×200 are randomly generated from these three image pairs. In the network testing, a pair of images is utilized, where the size of the MODIS image is 60×60, of the others is 3000×3000. Their center location is (95.78W, 29.85N).

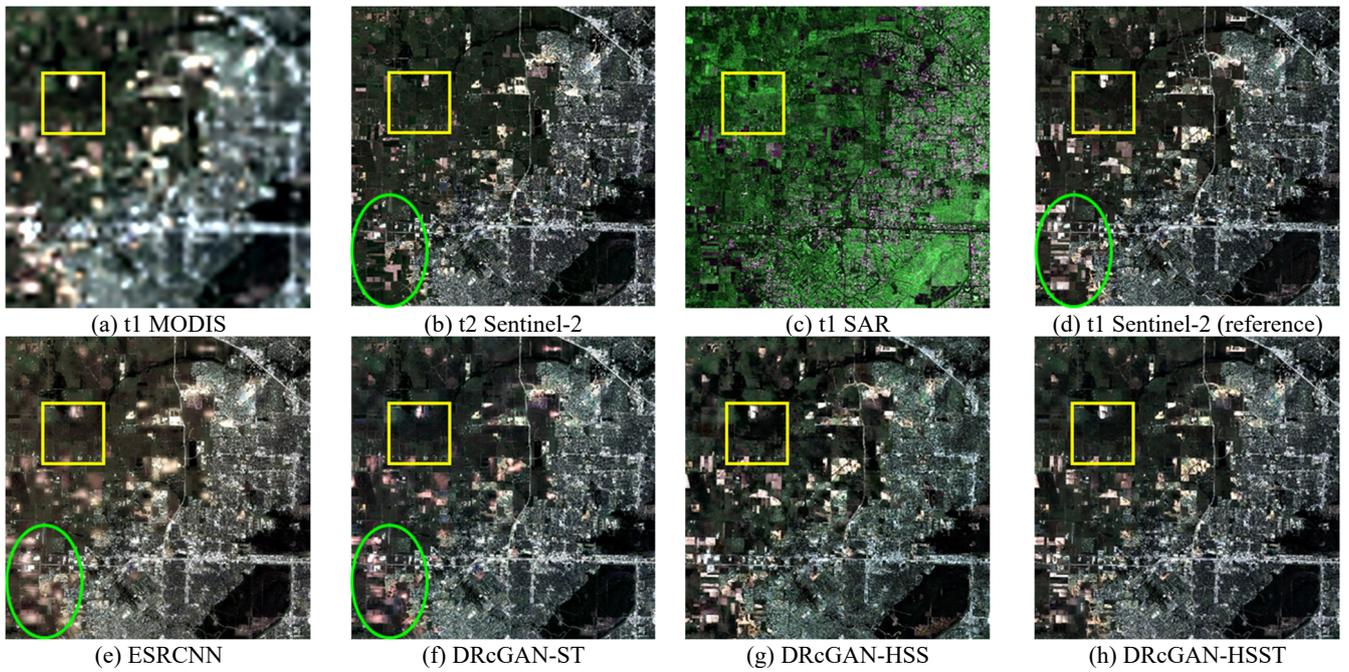
**Fig. 4** Fusion results of the spatial resolution improvement experiment

**3.2.2. Simulated experiments of spatial resolution improvement**

Fig. 4 displays the fusion results of the spatial resolution improvement experiment, where the optical images are displayed in the red-green-blue band combination, and the SAR image is displayed in the VV-VH-VV band combination. In the first row, the first three are observations, including the t1 MODIS image, the t2 Sentinel-2 image, and the t1 dual-polarization Sentinel-1 SAR image. The fourth is the t1 Sentinel-2 image used as the reference. The second row shows the fusion results of various approaches, where the first is the fusion result of the benchmark algorithm ESRCNN. The second is the fusion result of the proposed deep residual cycle GAN with the spatio-temporal fusion strategy of fusing the t1 MODIS image and the t2 Sentinel-2 image, which is written as DRcGAN-ST for brevity. The third is the fusion result of the proposed method with the heterogeneous spatio-spectral fusion strategy of fusing the t1 MODIS image and the t1 SAR image, which is written as DRcGAN-HSS for brevity. The fourth is the fusion result of the proposed method with the heterogeneous-integrated fusion strategy of fusing the t1 MODIS image, t1 SAR image, and t2 Sentinel-2 image, which is written as DRcGAN-HSST for brevity.

It is worth noting that lots of land cover changes happen between t2 and t1, as can be found from Sentinel-2 images at t1 and t2, for example, from the green ellipses in Figs. 4 (b) and (d). As shown in Fig. 4, in general, the four methods have improved the spatial information of the observed MODIS image. But comparing the fusion results with the reference, it can be found that the reference is dark green in the vegetation area, while the fusion result of ESRCNN is brown, showing some spectral distortion. In addition, focus on the two spatio-temporal fusion-based methods ESRCNN and DRcGAN-ST, in the areas with land cover changes, such as the green ellipse in Figs. 4 (e) and (f), ESRCNN has detected the land cover changes but fails to reconstruct them, resulting in blurring in these areas, while DRcGAN-ST shows spectral distortion in these areas. On the whole, results of DRcGAN-HSS and DRcGAN-HSST in Figs. 4 (g-h) seem closer to the reference in Fig. 4 (d).

To further analyze the effects of these methods, Fig. 5 displays the area inside the yellow rectangle in Fig. 4. In Fig. 4, the red rectangle points out one of the most visible land cover changes from t2 to t1, which is zoomed in the lower-right corner. Among the various methods, consistent with Fig. 4, ESRCNN shows the global spectral distortion, DRcGAN-ST shows local spectral distortion. Both of them perform poorly in the changed land covers, as shown in the white building in the zoomed areas in Fig. 5 (e-f). The reason is that neither the MODIS image at t1 nor the Sentinel-2 image at t2 contains the high-resolution information of the changed land covers. On the contrary, DRcGAN-HSS that uses the SAR image containing the high-resolution information of the changed land covers performs well in these areas, as can be seen in the zoomed areas in Figs. 5 (g). However, since heterogeneous information fusion is more difficult than homogeneous information fusion, DRcGAN-HSS has overall greater spatial distortion than DRcGAN-ST, as shown in the large cultivated land areas in the left of Fig. 5 (g). The DRcGAN-HSST method combines the advantages of homogeneous information transformation in DRcGAN-ST and reflecting land cover changes in DRcGAN-HSS. It not only performs well spatially and spectrally but also performs outstandingly in predicting land cover changes, whose fusion result is closest to the reference, as shown in Fig. 5 (h).

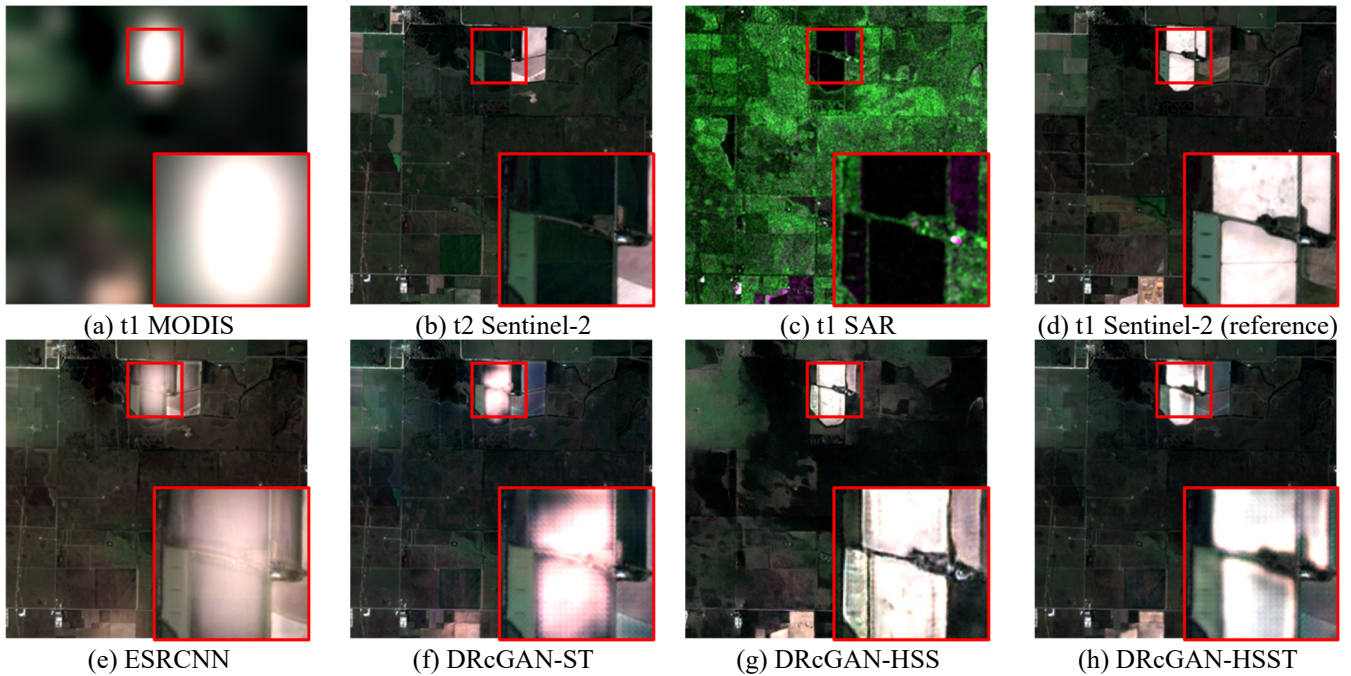

**Fig. 5** The sampled area of the spatial resolution improvement experiment.

Table II lists the quantitative evaluation results for the spatial resolution improvement experiments, where the best performance for each index is marked in bold. In table II, DRcGAN-HSS is the worst in all quality indices due to the difficulty in heterogeneous information fusion. DRcGAN-ST outperforms ESRCNN in all the quality indices, which shows the effectiveness of the proposed deep-residual-cycle-GAN. DRcGAN-HSST performs best in all quality indices, which verifies the superiority of the proposed heterogeneous-integrated fusion strategy.

Table II

QUANTITATIVE RESULTS OF SPATIAL RESOLUTION IMPROVEMENT EXPERIMENTS

| Algorithm | SAM | ERGAS | Q | PSNR | $SSIM_B$ | $SSIM_G$ | $SSIM_R$ | $SSIM_{AVG}$ |
|---|---|---|---|---|---|---|---|---|
| Ideal data | 0 | 0 | 1 | +∞ | 1 | 1 | 1 | 1 |
| DRcGAN-HSS | 7.8101 | 1.0268 | 0.3375 | 20.3921 | 0.6794 | 0.6437 | 0.5575 | 0.6269 |
| ESRCNN | 5.8961 | 0.8241 | 0.6712 | 22.3265 | 0.8460 | 0.8166 | 0.7335 | 0.7987 |
| DRcGAN-ST | 5.6597 | 0.7838 | 0.6775 | 22.8191 | 0.8592 | 0.8358 | 0.7546 | 0.8166 |
| **DRcGAN-HSST** | **5.2453** | **0.6697** | **0.7185** | **24.0525** | **0.8889** | **0.8703** | **0.8169** | **0.8587** |

## 3.3. Thick cloud removal experiments

### 3.3.1 Data sets for training and testing

Table III

DATA SETS OF THICK CLOUD REMOVAL EXPERIMENTS

| | Sensor | Time | Resolution | Size (train) | Size (test) | Cloudy (train/test: %) | Location (train/test) |
|---|---|---|---|---|---|---|---|
| **Cloud Removal** | Sentinel-2 MS | 2019-12-22 | 10m | 5830×10580×3 | 5830×400×3 | 19.72/23.15 | 88.44W, 41.96N/ 88.43W, 41.47N |
| | Sentinel-1 SAR | 2019-12-21 | 10m | 5830×10580×2 | 5830×400×2 | 0 | |
| | Sentinel-2 MS | 2019-06-10 | 10m | 5830×10580×3 | 5830×400×3 | 0 | |
| | Sentinel-2 MS | 2019-12-22 | 10m | 5830×10580×3 | 5830×400×3 | 0 | |

In the thick cloud removal experiments, we remove the thick clouds of Sentinel-2 images at the target time t1 with the help of the Sentinel-1 SAR image at t1 and the cloudless Sentinel-2 image at another time t2. Since we cannot simultaneously capture both cloudy and cloudless images of the same day, in the network training and simulated experiments, we synthesize the t1 cloudy Sentinel-2 image by adding a cloud mask to the observed t1 cloudless Sentinel-2 image. Then, the t1 cloudless Sentinel-2 observation works as label data of the network training and the reference of simulated experiments. Table III lists the datasets for the network training and testing. In table III, 2019-12-22 and 2019-12-21 are t1, and 2019-06-10 is t2. One image pair of size 5830×10580, whose center location is (88.44W, 41.96N), is used to generate the training patches. The cloud coverage of the synthesized cloudy Sentinel-2 image at t1 is 19.72%. In total, 8112 patches of size 128×128 are randomly generated. In the simulated experiments, a pair of images of size 5830×400, whose center location is (88.43W, 41.47N), is utilized to generate 22 small image pairs with the size of 256 × 256. The cloud coverage of the synthesized t1 cloudy Sentinel-2 image is 23.15%.

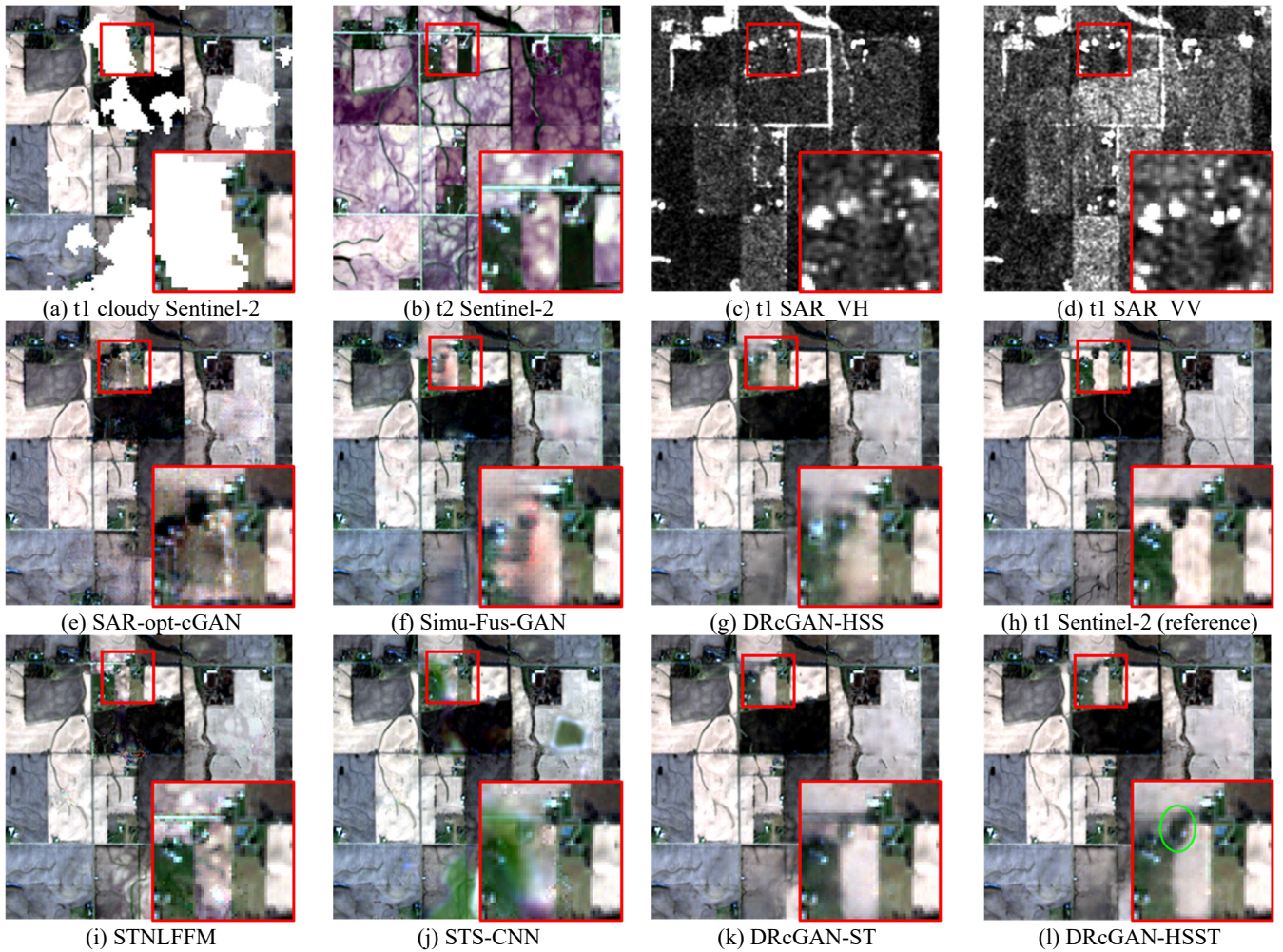

**Fig. 6** Results of the simulated thick cloud removal experiment

### 3.3.2 Simulated thick cloud removal experiments

A group of simulated experiments is selected to be displayed in Fig. 6 in the red-green-blue band combination, where the lower-right corner is a magnified display of the image inside the red rectangle. In Fig. 6, the first row displays the observations, including the t1 cloudy Sentinel-2 image, the t2 cloudless Sentinel-2 image, and the t1 dual-polarization Sentinel-1 SAR image. The second row shows the results of the heterogeneous fusion-based methods of fusing SAR and cloudy Sentinel-2 images at t1 and shows the reference. Among them, Fig. 6 (g) is the result of the proposed method of the heterogeneous spatio-spectral fusion strategy. The first three in the third row show the results of the spatio-temporal fusion-based methods of fusing the cloudy Sentinel-2 image at t1 and the cloudless Sentinel-2 image at t2. Among them, Fig. 6 (k) is the result of the proposed method of the spatio-temporal fusion strategy. The last one in the third row is the proposed method of the heterogeneous spatio-spectral-temporal integrated fusion strategy.

First, in the heterogeneous fusion-based methods, SAR-opt-cGAN has severe spatial distortion, Simu-Fus-GAN has obvious spectral distortion, as shown by the bare soil in the zoomed areas of Figs. 6 (e-f). The fusion result of the proposed DRcGAN-HSS is closer to the reference image, but it has some blurring spatially, as shown in the edges of the left white bail soil in the zoomed area of Fig. 6 (g). Then, for the spatio-temporal fusion-based method, spatially, the result of STNLFFM in the cloud coverage areas is close to that of the t2 Sentinel-2 image but is far from that of the reference image, as can be found in the zoomed areas in Figs. 6 (i). The result of STS-CNN shows obvious spectral distortion, as shown in the zoomed area in Fig. 6 (j). In contrast, the result of the proposed DRcGAN-ST is closer to the reference spatially and spectrally. Moreover, comparing the proposed three methods, DRcGAN-HSS shows some blurring in the zoomed area in Fig. 6 (g), DRcGAN-ST is better than DRcGAN-HSS but slightly worse than HDRcGAN-HSST, as can be seen by the vegetation in the green ellipse of the zoomed areas in Fig. 6 (l). The result of DRcGAN-HSST is closest to the reference, which shows the superiority of the heterogeneous-integrated strategy.

Table IV
QUANTITATIVE RESULTS OF SIMULATED THICK CLOUD REMOVAL EXPERIMENTS (22 GROUPS)

| Algorithm | SAM | Q | PSNR | $SSIM_B$ | $SSIM_G$ | $SSIM_R$ | $SSIM_{AVG}$ |
|---|---|---|---|---|---|---|---|
| Ideal data | 0 | 1 | +∞ | 1 | 1 | 1 | 1 |
| SAR-opt-cGAN | 3.3106 | 0.8277 | 24.5492 | 0.8261 | 0.8235 | 0.7781 | 0.8092 |
| Simu-Fus-GAN | 3.5359 | 0.8523 | 25.3729 | 0.8741 | 0.8532 | 0.8026 | 0.8433 |
| DRcGAN-HSS | 1.3511 | 0.8940 | 28.8842 | 0.9197 | 0.9015 | 0.8712 | 0.8975 |
| STNLFFM | 1.7946 | 0.8833 | 27.7753 | 0.9018 | 0.8807 | 0.8412 | 0.8746 |
| STS-CNN | 2.2743 | 0.8647 | 26.3756 | 0.8937 | 0.8801 | 0.8483 | 0.8740 |
| DRcGAN-ST | 1.3439 | 0.9120 | 29.7203 | 0.9282 | 0.9123 | 0.8853 | 0.9086 |
| DRcGAN-HSST | **1.2132** | **0.9200** | **30.2195** | **0.9131** | **0.9304** | **0.9169** | **0.8919** |

Table IV lists the quantitative evaluation results for the simulated thick cloud removal experiments with an average of 22 groups. In table IV, columns 2–4 list three heterogeneous fusion-based methods, where the best performance for each index is marked in blue. Columns 5-7 list three spatio-temporal fusion-based approaches, where the best performance is marked in red. Column 8 lists the heterogeneous-integrated fusion-based method. Among all the approaches in table IV, the best performance for each index is marked in bold. First, in the heterogeneous fusion-based methods, the proposed DRcGAN-HSS method outperforms the other two methods in all quality indices. It is also true in the spatio-temporal fusion-based methods in columns 5-7, where the proposed DRcGAN-ST performs better than the other two methods in all indices. The good performances of the proposed DRcGAN-HSS and DRcGAN-ST show the superiority of the proposed network. Furthermore, comparing the results of DRcGAN-HSS, DRcGAN-ST, and DRcGAN-HSST, DRcGAN-ST performs better than DRcGAN-HSS in all indices. DRcGAN-HSST performs the best in all indices, which shows the effectiveness of the heterogeneous-integrated fusion strategy.

### 3.4. Simultaneous thick cloud removal and spatial resolution improvement experiments
### 3.4.1 Data sets for training and testing

For the experiments of simultaneous thick cloud removal and spatial resolution improvement, we remove the thick clouds and improve the spatial resolution of the Landsat-8 image at t1. The auxiliary data include the Sentinel-1 SAR

Table V

DATA SETS OF THICK CLOUD REMOVAL AND SPATIAL RESOLUTION IMPROVEMENT EXPERIMENTS

| | Sensor | Time | Resolution | Size (train) | Size (test) | Cloudy (train/test: %) | Location (train/test) |
|---|---|---|---|---|---|---|---|
| **Thick Cloud Removal and Spatial Resolution Improvement** | Landsat-8 | 2017-10-15 | 30m | 1900×2794×3 | 240×2794×3 | 32.58/28.28 | 95.65W, 30.09N/ 95.32W, 30.10N |
| | Sentinel-1 SAR | 2017-10-28 | 10m | 5700×8382×2 | 720×8382×2 | 0 | |
| | Sentinel-2 MS | 2016-09-29 | 10m | 5700×8382×3 | 720×8382×3 | 0 | |
| | Sentinel-2 MS | 2017-10-29 | 10m | 5700×8382×3 | 720×8382×3 | 0 | |

image at t1 and the cloudless Sentinel-2 image at t2. In the network training and simulated experiments, the t1 cloudy Landsat-8 image is synthesized by adding a cloud mask to the t1 Sentinel-2 image that is spatially down-sampled to the resolution of the Landsat-8 image. Table V lists the training and testing datasets. In table V, 2016-09-29 is t2; and the other three are t1. One pair of images is utilized to generate the training patches, where the size of the Landsat-8 image is 1900×2794, the size of the others is 5700×8382. Their center location is (95.65W, 30.09N). The cloud coverage of the Landsat-8 image in network training is 32.58%. In total, 6336 patches of size 128×128 are randomly generated. In the simulated experiments, a pair of images is utilized, where the size of the Landsat-8 image is 240×2794, whose center location is (95.32W, 30.10N). The cloud coverage of the synthesized Landsat-8 image is 28.28%. They are utilized to generate 8 representative image pairs with the size of 256 × 256. In real-data experiments, the t1 cloudy Landsat-8 is captured by the Landsat-8 sensor, which has the same parameters as that in simulated experiments except that the cloud coverage is 34.51%.

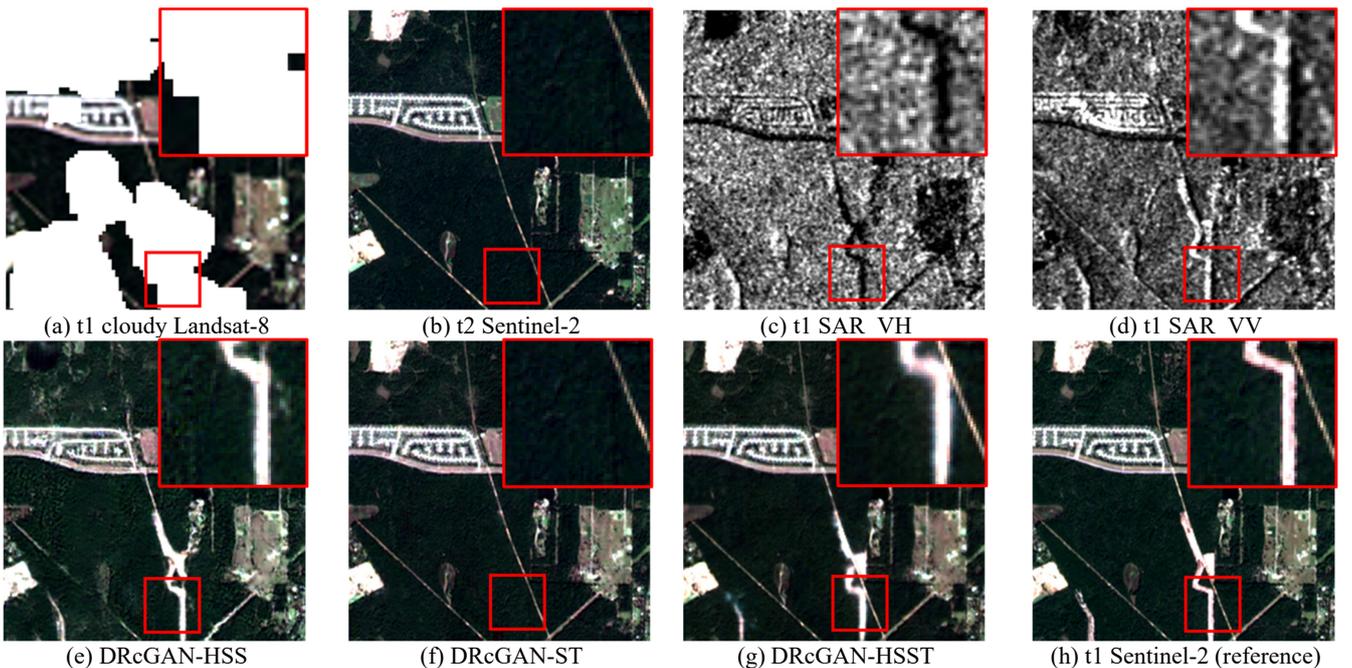

(a) t1 cloudy Landsat-8    (b) t2 Sentinel-2    (c) t1 SAR VH    (d) t1 SAR VV

(e) DRcGAN-HSS    (f) DRcGAN-ST    (g) DRcGAN-HSST    (h) t1 Sentinel-2 (reference)

**Fig. 7** Results of the simulated thick cloud removal and spatial resolution improvement experiments

### 3.4.2 Simulated experiments of thick cloud removal and spatial resolution improvement

Since there are few studies for the simultaneous thick cloud removal and spatial resolution improvement, in this section, we focus on comparing the proposed three methods: DRcGAN-HSS, DRcGAN-ST, and DRcGAN-HSST. A group of simulated experiments is selected to be displayed in Fig. 7 in the red-green-blue band combination, where the upper-right corner is a magnified display of the image inside the red rectangle. From the fusion results, it can be seen that all methods have effectively removed thick clouds and improved spatial structure information of the t1 cloudy Landsat-8 image in Fig. 7 (a). More in detail, as shown in the t2 Sentinel-2 image in Fig. 7 (b) and the reference in Fig. 7 (h), the red rectangle points out one of the most obvious land cover changes between t2 and t1, where the white building next to the road doesn't exist at t2, but it does at t1. Unfortunately, in the result of the DRcGAN-ST method using the t2 Sentinel-2 image as an auxiliary in Fig. 7 (f), the white building is not seen, which shows the inability of the spatio-temporal fusion strategy in predicting land cover changes. On the contrary, the t1 SAR image contains the information of the white building, as shown in Fig. 7 (c) and (d). Thus, DRcGAN-HSS using the SAR image as an auxiliary have effectively reconstructed the white building. However, due to the difference in imaging mechanism, the clear road next to the white building in zoomed area in Fig. 7 (h) is unclear in the SAR image in Figs. 7 (c-d), resulting in the invisible road in the fusion result, as shown in Fig. 7 (e). A similar situation can be seen in the white build of the upper left corner in Figs. 7 (e) and (h). The DRcGAN-HSST method integrates the characteristics of DRcGAN-HSS and DRcGAN-ST, makes full use of the complementary information of the t1 SAR image and the t2 Sentinel-2 image, and effectively reconstructs both the road and the white building in the zoomed area in Fig. 7 (g).

Table VI lists the quantitative evaluation results of the simulated cloud removal and spatial resolution improvement experiments with an average of 8 groups, where the best performance for each index is marked in bold. In table VI, the DRcGAN-HSS method performs the worst in all indices. The reason is that the transformation between heterogeneous information is more difficult than that between homogeneous information. The DRcGAN-HSST method is slightly better than the DRcGAN-ST method in all indices expected $SSIM_G$, which shows the superiority of the heterogeneous-integrated fusion strategy.

Table VI
QUANTITATIVE RESULTS OF SIMULATED THICK CLOUD REMOVAL AND SPATIAL RESOLUTION IMPROVEMENT EXPERIMENTS (8 GROUPS)

| Algorithm | SAM | ERGAS | Q | PSNR | $SSIM_B$ | $SSIM_G$ | $SSIM_R$ | $SSIM_{AVG}$ |
|---|---|---|---|---|---|---|---|---|
| Ideal data | 0 | 0 | 1 | +∞ | 1 | 1 | 1 | 1 |
| DRcGAN-HSS | 5.8573 | 19.8796 | 0.5741 | 24.8477 | 0.8285 | 0.7845 | 0.7327 | 0.7819 |
| DRcGAN-ST | 4.0801 | 13.4593 | 0.8375 | 28.2400 | 0.9439 | **0.9328** | 0.8992 | 0.9253 |
| DRcGAN-HSST | **4.0198** | **12.1935** | **0.8396** | **29.0881** | **0.9457** | 0.9325 | **0.9010** | **0.9264** |

### 3.4.3 Real-data experiments of thick cloud removal and spatial resolution improvement

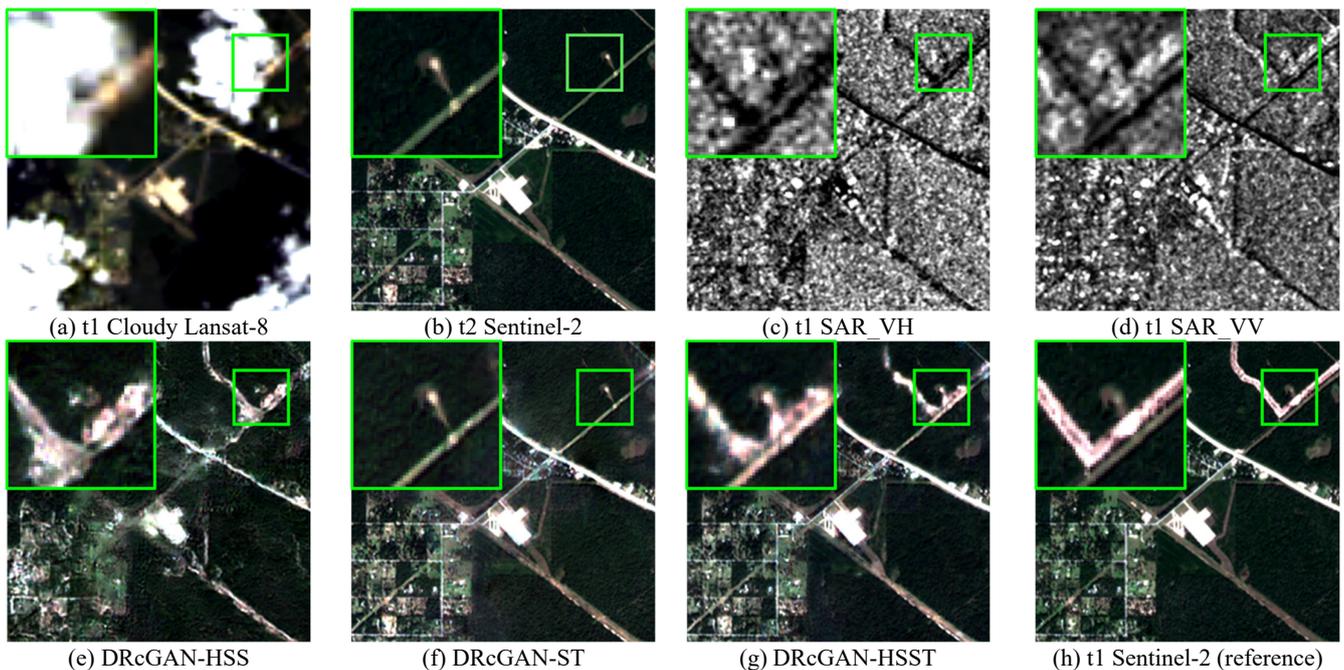

(a) t1 Cloudy Lansat-8  (b) t2 Sentinel-2  (c) t1 SAR_VH  (d) t1 SAR_VV
(e) DRcGAN-HSS  (f) DRcGAN-ST  (g) DRcGAN-HSST  (h) t1 Sentinel-2 (reference)

**Fig. 8** Results of the real-data experiment for simultaneous thick cloud removal and spatial resolution improvement

To further compare the proposed methods, a group of real-data experiments is selected to be displayed in Fig. 8 in the red-green-blue band combination, where the upper-left corner is a magnified display of the image inside the green rectangle. In the real-data experiment, the cloudy image is not synthetic, which is directly obtained from the Landsat-8 sensor. As shown in Fig. 8, all methods remove the thick clouds in the cloudy Landsat-8 image in Fig. 8 (a) without a trace and effectively enhance its spatial structure information. However, the result of DRcGAN-HSS has severe global spatial distortion, as shown in the lower-left corner of Fig. 8 (e). DRcGAN-ST fails to reconstruct the changed land covers between t2 and t1, as can be found by comparing the zoomed areas in Figs. 8 (f) and (h). DRcGAN-HSST makes full use of the complementary of t1 SAR and t2 Sentinel-2 images. It not only obtains an overall good result but also successfully reconstructs the changed land covers.

## 4. Conclusions and future prospects

In this paper, a deep residual cycle GAN-based heterogeneous-integrated fusion framework is proposed. The proposed network considers the imaging degradation process and includes a forward fusion part and a backward feedback part. For the first time, a heterogeneous-integrated fusion framework is proposed to simultaneously fuse the complementary spatial, spectral, and temporal information between multi-sources heterogeneous observations. In addition, the proposed method provides a unified framework that can deal with various tasks, including heterogeneous spatio-spectral fusion, spatio-temporal fusion, and heterogeneous spatio-spectral-temporal integrated fusion. Three types of experiments are carried out on two difficult problems of land cover changes and thick cloud removal. Datasets from multiple remote sensing satellites, including MODIS, Landsat-8, Sentinel-1, Sentinel-2, are utilized in experiments. Both qualitative and quantitative evaluations have demonstrated the effectiveness of the proposed network and the superiority of the proposed heterogeneous-integrated fusion strategy. The proposed heterogeneous-integrated fusion strategy can combine the advantages of the heterogeneous spatio-spectral fusion strategy and the spatio-temporal fusion strategy to make up for their shortcomings.

The proposed methods have experimented on only the red, green, and blue bands of MS images. In the future, we will extend them to more bands of MS images and even HS images. In addition, in the backward degradation part of the proposed network, we only explicitly utilize the spatial degradation between the t1 LR MS and the t1 HR MS images. In our future work, we will further explore the heterogeneous spectral relationship model between the t1 SAR and t1 HR MS images, the temporal relationship model between t2 HR MS and t1 HR MS images, and embed them into the network design. Last but not least, the proposed methods focus on the image reconstructions of a single target time. It will be of great significance to extend the heterogeneous-integrated fusion to the time-series image reconstruction.